\documentclass[a4paper,UKenglish]{lipics}
\usepackage{clrscode3e}
\usepackage{varwidth}
\usepackage{latexsym}
\usepackage[boxed,linesnumbered]{algorithm2e}
\usepackage{cite,booktabs}
\usepackage{tikz}
\usepackage{color}
\usepackage{comment}
\usepackage{enumitem}
\usepackage{tabto}

\usepackage{calligra}
\usepackage[T1]{fontenc}

\usetikzlibrary{shapes}
\usetikzlibrary{calc}

\DontPrintSemicolon

\newcommand{\NP}{\ensuremath{\mathsf{NP}}}
\newcommand{\NPI}{\ensuremath{\mathsf{NP}}-Intermediate}
\newcommand{\SAT}{\ensuremath{\mathsf{SAT}}}
\renewcommand{\P}{\ensuremath{\mathsf{P}}}

\newcommand{\q}{\mathcal Q}
\newcommand{\qnext}{\mathcal Q_{next}}

\newtheorem{Fact}{Fact}

\newcommand{\ceil}[1]{\lceil #1 \rceil}
\newcommand{\floor}[1]{\lfloor #1 \rfloor}

\newcommand{\logclique}{L{\small OG}C{\small LIQUE}}

\theoremstyle{plain}

\makeatletter
\newenvironment{proofof}[1]{\par
  \pushQED{\qed}%
  \normalfont \topsep6\p@\@plus6\p@\relax
  \trivlist
  \item[\hskip\labelsep
        \bfseries
    Proof of #1\@addpunct{.}]\ignorespaces
}{%
  \popQED\endtrivlist\@endpefalse
}
\makeatother

\newcommand{\rf}{Restricted-$f$-Factor}

\newcommand{\polylog}[1]{\textnormal{polylog}($#1$)}
\title{A  Classification of Connected $f$-factor Problems inside \NP \footnote{Supported by the Indo-German Max Planck Center for Computer Science grant for the year 2013-2014 in the area of Algorithms and Complexity}}

\author[1]{N.S. Narayanaswamy}
\author[2]{C.S. Rahul}
\affil[1]{Indian Institute of Technology Madras, Chennai, India\\
  \texttt{swamy@cse.iitm.ac.in}}
\affil[2]{Indian Institute of Technology Madras, Chennai, India\\
  \texttt{rahulcs@cse.iitm.ac.in}}
\authorrunning{Narayanaswamy and Rahul} 

\Copyright{N.S. Narayanaswamy and C.S. Rahul}

\subjclass{"F.1.3 Complexity Measures and Classes", "G.2.2 Graph Theory}
\keywords{$f$-factors, Connected $f$-factors, \NP-Complete, \NPI, \P-time}

\serieslogo{}
\volumeinfo
  {Billy Editor and Bill Editors}
  {2}
  {Conference title on which this volume is based on}
  {1}
  {1}
  {1}
\EventShortName{}
\DOI{10.4230/LIPIcs.xxx.yyy.p}

\begin{document}
\maketitle
\begin{abstract}
Given an undirected graph $G=(V,E)$ with $|V|=n$, and a function $f:V\rightarrow \mathbb{N}$, we consider the problem of finding a connected $f$-factor in $G$. This problem is \NP-Complete when $f(v)\geq n^{\epsilon}$ for every $v$ in $V$ and a contant $\epsilon>0$. We design an algorithm to check for the existence of a connected $f$-factor,  for the case where  $f(v)\geq n/g(n)$, for all $v$ in $ V$ and $g(n)$ is polylogarithmic in $n$. The running time of our algorithm is $\tilde{\mathcal O}(n^{2g(n)})$.  As a consequence of this algorithm, we conclude that the complexity of connected $f$-factor for the case we consider is unlikely to be \NP-Complete unless the Exponential Time Hypothesis (ETH) is false.   Secondly, under the ETH  assumption, we show that the problem is also unlikely to be in \P ~for $g(n)$ in $\mathcal O((\log{n})^{1+\epsilon})$ for any constant $\epsilon > 0$.  These results show that for each $\epsilon > 0$ and $g(n)$ in $\mathcal O((\log{n})^{1+\epsilon})$, connected $f$-factor problem for $f(v) \geq n/g(n)$ is in \NPI\  unless the ETH is false.   Further, for any constant $c > 0$, when $g(n)=c$, our algorithm for connected $f$-factor runs in polynomial time. Finally we extend our algorithm to compute a minimum weight connected $f$-factor in edge weighted graphs in the same asymptotic time bounds.  
\end{abstract}
\section{Introduction}
\noindent
Let $G=(V,E)$ be an undirected graph with $n$ vertices and $f:V\rightarrow \mathbb N$ be a function.  An $f$-factor \cite{DW00} of $G$ is a spanning subgraph $H$ such that $d_H(v)=f(v)$, for each $v$ in $V$. The problem of deciding whether a given graph $G$ has an $f$-factor is a well studied problem over many years \cite{An85,tutte1952factors,Tokuda1999293,cornuejols1988general,iida1991ore,PM07} and the problem is shown to be polynomial time solvable by Tutte \cite{WT54}. When edges have weights, a simple modification to Tutte's reduction solves the minimum weighted $f$-factor problem. \\
A connected $f$-factor is an $f$-factor which is connected. For the case when $f(v) = 2$ for all $v$ in $V$, a connected $f$-factor is a Hamiltonian cycle \cite{DW00} and is \NP-Complete to decide.   In fact, Cheah and Corneil \cite{CC90} showed that the  connected $f$-factor problem  is \NP-Complete where $f(v) = d$ for each $v$ in $V$ and an integer constant $d>1$. For $f(v)\geq\lceil\frac n2 \rceil$ for every $v$ in $V$, deciding a connected $f$-factor is same as deciding whether there exists an $f$-factor or not. This is because in this case, any $f$-factor turns out to be connected, due to Ore~\cite{OR60} and Dirac~\cite{DI52}.\\
 {\bf Past Work on Connected Factors.} There has been an extensive study on connected $[a,b]$-factors in the literature over the past twenty years.  An $[a,b]$-factor is a subgraph $H$ of a graph $G$ such $a \leq d_H(v) \leq b$,  for each $v $ in $ V$.  There are many results on sufficiency conditions for a graph to have a connected $[a,b]$-factor.  For example, when $\delta(G) \geq \frac{n}{2}$ the Graph is Hamiltonian, due to Ore~\cite{OR60} and Dirac~\cite{DI52}.  Also, if the sum of degrees of every pair of non-adjacent vertices is at least  $n-1$, then the graph has a Hamilton path, and this is a connected $[1,2]$-factor.   Similarly, by relating the size of the maximum independent set and the vertex connectivity of a graph, there are sufficiency conditions for the existence of connected $[a,b]$-factors.  The survey article by Kouider and Vestergaard \cite{KV05} and Plummer \cite{PM07} present more results on connected $f$-factors.\\
{\bf Our Work.}
To the best of our knowledge, our study is the first of this kind in the area of connected factors.  We are motivated by this line of study with an aim to classify functions $f$ for which the connected $f$-factor problem is polynomial time solvable and those for which the problem is  \NP-Complete.  In particular, our interest is to obtain a dichotomy for connected $f$-factor problem based on $f$.  To conceptualize the nature of $f$, $f(v)$ is taken to be at least  $n/g(n)$ for each $v$ in $V$ where $g(n)$ is a function in $o(n)$.   In \cite{BN13} we have shown that the problem is \NP-Complete when $g(n)$ is $n^{1-\epsilon}$ for any constant $\epsilon$ between 0 and 1.  In recent work ~\cite{NR15}, we showed that the problem is polynomial time solvable if $g(n)=3$.
While connected $[a,b]$-factors are studied extensively from the point of view of identifying sufficient conditions, our work is on understanding how the computational complexity of connected $f$-factor problem vary with $f$.
We summarize our results as follows:
\begin{enumerate}
\item An algorithm running in time $\tilde{\mathcal O}(n^{2g(n)})$ for deciding the existence of connected $f$-factor in a graph $G$ where $f(v)\geq n/g(n)$ for each $v$ in $V$ and $g(n)$ is in $\mathcal O(\polylog n)$. Clearly, the algorithm takes  polynomial time when $g(n)$ is a constant and quasi-polynomial time when $g(n)$ is polylogarithmic in $n$.   It is interesting that connected $\frac{n}{c}$-factor problem is polynomial time solvable for any constant $c$, as this refines the class of functions $f$ for which the connected $f$-factor problem is \NP-Complete: connected $d$-factor is \NP-Complete for each constant $d$, as shown by Cheah and Corneil in \cite{CC90}. 
\item A \emph{refined} characterization of graphs having connected $f$-factor where $f(v)\geq n/g(n)$ for every $v$ in $V$ and $g(n)$ is in $\mathcal O(\polylog n)$.
\item An extension of the above mentioned algorithm to solve the minimum weighted connected $f$-factor problem where $f(v)\geq n/g(n)$ for every $v$ in $V$ and $g(n)$ is in $\mathcal O(\polylog n)$, without increasing the asymptotic running time.
\item  Connected $f$-factor problem for $f(v)\geq n/g(n)$ for every $v$ in $V$ and $g(n)$ is in $\mathcal O((\log n)^{1+\epsilon})$  for any constant $\epsilon>0$, is in \NPI\  under the ETH\cite{IR99}. Thus, this infinite class of problems parameterized by $\epsilon$ is similar in complexity to the \logclique\cite{PV15} problem where the goal is to decide whether there exists a clique of size $\log n$ in an $n$-vertex graph.  
\end{enumerate}
As a consequence of this work, we have a better refined understanding of computational complexity of the connected $f$-factor problem based on the nature of $f$. The main technique in this work 
is a natural way of converting one $f$-factor to another by exchanging a set of edges. This is formalized using the notion of \emph{Alternating Circuit}s that we use extensively in this work. We believe that these techniques for enforcing connectedness along with the results of Tutte \cite{WT54} for finding $f$-factors plays an important role in understanding the nature of the connected $f$-factor problem for different classes of functions $f$.
\section{Preliminaries}
\subsection{Definitions and Notations}
\noindent
We use standard definitions and notations from West \cite{DW00}.  $G=(V,E)$  represents an  undirected graph on $n$ vertices,  $d_G(v)$ denotes the degree of a vertex $v$ in a graph $G$ and 
$N(v)$ denotes the open neighborhood of a vertex $v$. $g(n)$ is in $\mathcal O(\polylog{n})$ and $f$ is a function whose domain is the vertex set of $G$ and range is the set $\{\lceil n/g(n) \rceil,\ldots,n-1 \}$.  
Given two subgraphs $G_1$ and $G_2$ of a graph $G$, we use the basic definitions of binary operations $G_1\cap G_2$, $G_1\cup G_2$ to be  subgraphs obtained by the vertex and edge set  intersection and union operations respectively. We define the symmetric difference  $G_1\bigtriangleup G_2$ between two spanning subgraphs $G_1$ and $G_2$ of $G$ to be the spanning subgraph whose edge set is $E(G_1)\bigtriangleup E(G_2)$.  Further, the concepts of 
 circuit,  decomposition of a graph $G$, the subgraph of $G$ induced by $S \subseteq V$ denoted by $G[S]$  are standard.   
We use $w(e)$ to represent weight of an edge $e$ in a weighted graph and $w(G)$ to denote the sum of weights of edges in $G$. \\
Given a partition $\mathcal Q=\{Q_1,Q_2,\dots,Q_r\}$ of the vertex set of $G$, a graph $G/\mathcal Q$ is constructed as follows: The vertex set of $G/\mathcal Q$ is $\mathcal Q$. Corresponding to each edge $(u,v) $ in $ G$ where $u $ in $ Q_i$, $v$  in $ Q_j$, $i \neq j$, there exists an edge $(Q_i,Q_j)$ in $G/\mathcal Q$. $G/\mathcal Q$ is a multigraph without loops. For a spanning subgraph $G'$ of $G$, we say $G'$ \emph{connects} a partition $\mathcal Q$ if $G'/\mathcal Q$ is connected.  A refinement $\mathcal Q'$ of a  partition $\mathcal Q$ is a partition of $V$ where each part $Q'$ in $\mathcal Q'$ is a subset of some part $Q$ in $\mathcal Q$. This concept of partition refinement is from Kaiser \cite{TK12}. Whenever we say a spanning tree of $G/\mathcal Q$, we refer to a spanning subgraph $T$ of  $G$ having $|\mathcal Q|$-1 edges that connects $\mathcal Q$. 

\noindent
\subsection{Colored Graphs and Alternating Circuits}
A colored graph $G$ is one in which each edge is assigned a color from the set $\{red,blue\}$.
 In a colored graph $G$, we use $R$ and $B$ to denote subgraphs of $G$ whose edges are the set of red edges ($E(R)$) and blue edges ($E(B)$) of $G$, respectively, and $V(R)=V(B)=V(G)$. We use this coloring in our algorithm to distinguish between edge sets of two distinct $f$-factors of the same graph $G$. A main computation step in our algorithm is to consider the symmetric difference between edge sets of two distinct $f$-factors and perform a sequence of edge exchanges preserving the degree of each vertex. The following definition is  used extensively in our algorithm.
 \begin{definition}
A subgraph $S$ of a colored graph $G$ is an {alternating circuit} if $S$ is a circuit, and there exists an Eulerian tour of $S$ in which every pair of consecutive edges are of different colors. 
\end{definition}
\noindent
Clearly, an alternating circuit has an even number of edges and is connected. Further, $d_{R}(v)=d_{B}(v)$ for each $v$ in $S$. We define a \emph{minimal} alternating circuit $S$ to be an alternating circuit where each vertex $v$ in $S$ has at most two  red edges and two  blue edges incident on it.
\begin{definition}
A spanning subgraph $S$ of $G$ is defined to be a \emph{switch} on another spanning subgraph $H$ of $G$ if we could color edges in  $S\cap H$ with color red and those in $S\setminus H$ with color blue such that each component in $S$ is an alternating circuit.
\end{definition}
\begin{definition}
For an $S$ which is a switch on $H$,  we define Switching($H$,$S$) to be a subgraph $G'$ of $G$ obtained by removing all edges in $S\cap H$ from $H$ and adding all the edges in $S\setminus H$ to $H$.  
\end{definition}
\noindent
Whenever the operation Switching($H$,$S$) is used, $S$ is assumed to be a switch on $H$. Finally the  {\em weight of an alternating circuit $S$}, denoted by $W(S)$, is $w(B)-w(R)$.  This will be used along with switching operation. If $G'=$Switching($H$,$S$), then it implies that $w(G')=w(H)+W(S)$.  The weight of a switch $S$ is also similarly defined to be $w(S\setminus H)-w(S\cap H)$. In our arguments we reason about an $f$-factor obtained by switching a sequence of alternating circuits, and for this we introduce the following notation.  Let ${\mathcal S}$ be a set of edge disjoint alternating circuits each of which is a switch on $H$.  Let $S' = \displaystyle \cup_{S \in {\mathcal S}} S$.   Then the operation Switching($H$, ${\mathcal S}$) is  the $f$-factor that results from  Switching($H$,$S'$). 

\noindent
 Unless otherwise mentioned, $g(n)$ is a function in $\mathcal O($poly$\log(n))$. We justify why this choice of $g(n)$ is crucial for our
 analysis in Lemma \ref{lem:whypolylog}.  $f$ is a function $f:V \rightarrow \mathbb{N}$ such that $f(v) \geq \ceil{{n}/{g(n)}}$, for each $v $ in $ V$ where $n=|V|$. A consequent fact is that, if $H$ is an $f$-factor of $G$,  then the number of components in $H$ is at most $g(n)-1$.
We use two crucial subroutines from the literature-
 {\bf Tutte's-Reduction($G$,$f$)} is a subroutine which outputs an $f$-factor of $G$(if one exists) using the reduction in \cite[example 3.3.12]{DW00}.  {\bf Modified-Tutte's-Reduction($G$,$f$)} is an extension of Tutte's-Reduction($G$,$f$), which computes a minimum weighted $f$-factor of the input weighted graph $G$ by reducing it to the problem of finding a minimum weighted perfect matching \cite{EJ65,JE65}. We assume that both the above subroutines return  empty graphs if they fail to an compute $f$-factor.

\section{Outline of the Algorithm and a refined Characterization}
\label{sec:overview}

\noindent
The following is a natural characterization of graphs that have a connected $f$-factor, and it is almost a restatement of the definition of a connected $f$-factor.
\begin{theorem}
Let $G$ be an undirected graph  and $f$ be a function $f:V\rightarrow \mathbb N$.  $G$ has a connected $f$-factor if and only if for each partition $\mathcal Q$ of the vertex set $V$, there exists an $f$-factor $H$ of $G$ that connects $\mathcal Q$. 
\label{thm:characterization}
\end{theorem}
\noindent
The forward direction of the proof is the observation that a connected $f$-factor connects any partition of the vertex set. The converse is proved by applying the hypothesis to the partition  $\mathcal Q = \{\{v_1\},\{v_2\},\ldots,\{v_n\}\}$. Theorem \ref{thm:characterization} sets up the foundation of our algorithm outlined below. 

\noindent
{\bf Outline of the search for connected $f$-factors:}  Here we set up the template to search for a connected $f$-factor in an input graph $G$ based on Theorem \ref{thm:characterization}.  The details are in Algorithm \ref{Algm:1} in section \ref{sec:con-f-factor}. 
Our algorithm constructs a {\em maximal } sequence of pairs $(H_0,\mathcal Q_0), (H_1,\mathcal Q_1),\ldots,(H_k,\mathcal Q_k)$ satisfying the following properties:
\begin{enumerate}
\item Each $\mathcal Q_i, 0 \leq i \leq k $ is a partition of the vertex set $V$, and $\mathcal Q_0 =\{V\}$.
\item  Each $H_i, 0 \leq i \leq k$ is an $f$-factor of $G$, and $H_i$ connects $\mathcal Q_i$. 
\item {For each $ 1 \leq i \leq k$, Each $\mathcal Q_i$ is a refinement of $\mathcal Q_{i-1}$ satisfying the following: \begin{enumerate}
\item Each part in $\mathcal Q_i$ is a maximal component in $H_{i-1}[Y]$ for some $Y$ in $\mathcal Q_{i-1}$. 
\item $\mathcal Q_i \neq \mathcal Q_{i-1}$ and hence $|\mathcal Q_i|>|\mathcal Q_{i-1}|$  for $1 \leq i \leq k$.
\end{enumerate}}
\end{enumerate}
The meaning of maximality of the sequence is that the sequence we consider is not a prefix of a longer sequence satisfying the 3 conditions listed above.  
Since $\mathcal Q_i$ is a refinement of $\mathcal Q_{i-1}$, it follows that $k$ can be at most $n$. The following is an interesting and useful fact.
\begin{Fact}
Let $H$ be an $f$-factor of $G$ and let $\mathcal Q$ be a partitioning of the vertex set $V$. If $H/\mathcal Q$ is connected and $H[Q]$ is connected for each $Q$ in $\mathcal Q$, then $H$ is a connected $f$-factor.
\label{fac:connect-parts}
\end{Fact}
%
\noindent
If a refinement of $\mathcal Q_{k}$ satisfies conditions 1 and 3(a), is same as $\mathcal Q_k$, then it follows from Fact \ref{fac:connect-parts}, that $H_k$ is a connected $f$-factor of $G$. On the other hand, if $H_k$ is not a connected $f$-factor, then {\em any} refinement of $\mathcal Q_k$ has the property that there is no $f$-factor of $G$ that connects it.  Otherwise, if there was some refinement that can be connected by some $f$-factor of $G$, there would be  a violation to the maximality of the sequence. In this case, from Theorem \ref{thm:characterization} we can conclude that $G$ does not have a connected $f$-factor.
\noindent
The algorithm we design directly evolves from Theorem \ref{thm:characterization} and it  essentially computes a maximal sequence $(H_i, \mathcal Q_i), 0 \leq i \leq k$ of pairs satisfying the conditions outlined above. The algorithm initializes $\mathcal Q_0={V(G)}$, and $H_0$ to an arbitrary $f$-factor of $G$. The computation of the next pair in sequence is done by the subroutine Restricted-$f$-Factor($H_{i-1}$,$\mathcal Q_{i-1}$) (described later in Section \ref{sec:con-f-factor}) which computes $\mathcal Q_i$ from $\mathcal Q_{i-1}$ and $H_{i-1}$. Further the algorithm performs a computation of $H_i$ from $H_{i-1}$ and $\mathcal Q_{i}$. The procedure Restricted-$f$-Factor($H_{i-1}$,$\mathcal Q_{i-1}$) mainly involves three computational subtasks for each $1 \leq i \leq k$.
\begin{enumerate}
\item Computes $\mathcal Q_i$ from $\mathcal Q_{i-1}$ by considering the subgraph $H_{i-1}[Q]$ and finds components in $H_{i-1}[Q]$ for each $Q$ in $\mathcal Q_{i-1}$.  These components are all the parts in $\mathcal Q_i$.
 \item A set of steps referred to as Partition-Connector($\mathcal Q_i$) in the recursive procedure Restricted-$f$-Factor($H_{i-1}$,$\mathcal Q_{i-1}$) does the following: takes the partition $\mathcal Q_{i}$ as input and returns an $f$-factor $H$ such that $H/\mathcal Q_{i}$ is connected. This is achieved as follows: for each spanning tree $T$ of $G/\mathcal Q_{i}$(recall definition), the procedure Partition-Connector($\mathcal Q_{i}$) checks for an $f$-factor $H'_i$ containing $E(T)$. If there exists one, then it returns $H'_i$ and
an empty graph otherwise. The correctness of procedure Partition-Connector($\q_i$) uses Lemma \ref{lem:enforcedg}.
\item A sequence of steps referred to as Next-Factor($H_{i-1}$,$H'_i$,$\mathcal Q_i$) in procedure Restricted-$f$-Factor($H_{i-1}$,$\mathcal Q_{i-1}$) computes $H_{i}$ from $H_{i-1}$, $H'_i$ and $\mathcal Q_i$ where $H'_i$ is an $f$-factor that connects $\mathcal Q_i$. To compute $H_i$ we color edges in $H_{i-1}$ with color red and those in $H'_i$ with color blue. Then we consider the symmetric difference $S$ between $H_{i-1}$ and $H'_i$. Further we pick a set $\mathcal S$ of at most $|\mathcal Q_i|-1$ minimal alternating circuits in $S$ such that  Switching($H_{i-1}$,$\mathcal{S}$) connects $\mathcal Q_i$ and this is the required $H_i$. The correctness proof of the computation of $H_i$ from $H_{i-1}$ and $H'_i$ follows from Lemmas \ref{lem:minalterswitchlemma} and \ref{lem:min-ac-set}. 
\end{enumerate}
Using the above three subtasks in Restricted-$f$-Factor($H_{i-1}$,$\mathcal Q_{i-1}$), we ensure that at least $f(v)-2(g(n)-1)$ edges incident on $v$ in $H_{i-1}$ is present  in $H_{i}$ for each $v$ in $V$. We use this constructive computation of $H_{i}$ from $H_{i-1}$ to place a lower bound the size of the  smallest part in $\mathcal Q_i$ for every $i<k$. This gives an upper bound on the number of parts in $\mathcal Q_i$ and this consequently bounds the  number of recursive calls made to Restricted-$f$-Factor($H_{i-1}$,$\mathcal Q_{i-1}$), as $|\mathcal Q_i|>\mathcal |Q_{i-1}|$. These bounds are presented in Lemmas \ref{lem:partsizebound} and \ref{lem:noofcalls}. 
Based on our algorithm, we come up with a refined characterization for graphs having a connected $f$-factor, which is proved in Section \ref{sec:con-f-factor}.
\begin{theorem}
\label{thm:ref-characterization-Algm1}
Let $G$ be a graph and $f$ be a function where $f(v)\geq n/g(n)$ for every $v$ in $V$ and  $g(n)$ is a function in $\mathcal O(\polylog{n})$. $G$ has a connected $f$-factor if and only if for each partition $\mathcal Q$ of size at most $g(n)$, there exists an $f$-factor that connects $\mathcal Q$.  Further, Algorithm \ref{Algm:1} decides the existence of a connected $f$-factor in an input graph $G$ in time $\tilde{\mathcal O}(n^{2g(n)})$.
\end{theorem}
\section{A spectrum of $f$ for which Connected $f$-factor is in \NP-Intermediate under ETH}
\label{sec:npi}
Let $\epsilon > 0$ be a fixed number, and let $g(n)$ is in  $\mathcal O((\log  n)^{1+\epsilon})$. We consider the case when $f(v) \geq n/g(n)$ for all $v$ in $V$.  In this section we show that for each such $f$, the  connected $f$-factor problem is in \NPI\   under the ETH.   \\
\textbf{Connected $n/\mathcal O(\polylog{n})$-factor is unlikely to be \NP-Complete.} Assume  that the connected $f$-factor problem is \NP-Complete for some $f$ satisfying the condition mentioned above.
  This implies there exists a polynomial time reduction from 3-\SAT\  to the connected $f$-factor problem where the reduction algorithm outputs the graph $G$ and the function $f$ on the vertex set, both of which are polynomial in size of the instance of 3-\SAT. Further, we reiterate, $f$ satisfies the condition outlined above.
  From this reduction and the guarantee on our Algorithm in Theorem \ref{thm:ref-characterization-Algm1} for connected $f$-factor, it follows that we have an algorithm which decides 3-\SAT\ that runs in time $\tilde{\mathcal O}(n^{\textnormal{polylog}(n)})$, and  this is impossible under the ETH. The following lemma states this observation.
\begin{lemma}
The connected $f$-factor problem where $f(v)\geq n/g(n)$  for every $v$ in $V$ and $g(n)$ in $\mathcal O((\log  n)^{1+\epsilon})$  is not \NP-Complete for any $\epsilon \geq 0$ unless the ETH is false.  
\label{lem:non-npc}
\end{lemma}
\noindent
\textbf{Connected $(n/\mathcal O((\log n)^{1+\epsilon})$-factor is unlikely to be in \P.}
Here we assume that $f(v) \geq n/g(n)$ where $g(n)$ is in $\mathcal O((\log{n})^{1+\epsilon})$ for some $\epsilon > 0$.  Note that here $\epsilon > 0$ and in claim that it is unlikely to be \NP-Complete, we had assumed that $\epsilon \geq 0$.   
We now present a {\em sub-exponential} time reduction  $\textsf R$ from  the Hamiltonian cycle problem to the connected $f$-factor problem. The reduction algorithm takes $G$ on $N$ vertices and $\epsilon>0$ as input and outputs a set $\mathcal G$ of $\mathcal (|V(G)|^3)$ pairs.  Each pair in $\mathcal G$ is of the form $(G',f)$ where $f$ is a function and $G'$ is a graph. The set $\mathcal G$ satisfies the following:
 \begin{enumerate}
  \item For each pair $(G',f)$ in $\mathcal G$, $G'$ is a graph having $n$ vertices and $f(v)\geq n/\log^{1+\epsilon}n$ for every $v$ in $G'$.
  \item $G$ has a Hamiltonian cycle if and only if there exists a pair $(G',f)$ in $\mathcal G$ such that $G'$ has a connected $f$-factor. 
  \item For each $G'$ output by $\textsf R$, the number of edges in $G'$ is  $2^{o(N)}$ where $N$ is the number of vertices in $G$.
 \end{enumerate}
\noindent
The reduction $\textsf R$ is as follows-
\begin{enumerate}
 \item Compute $n=\ceil{2^{N^{1/(1+\epsilon)}}}$ where $N=|V(G)|$.
 \item Construct an empty graph $G'$ containing $n$ vertices.
 \item Define a partition $\mathcal Q$ of vertices in $G'$ where  each part contains at least $\ceil{\frac{n}{\log^{1+\epsilon}n}}+1$ vertices and $|\mathcal Q|=N-2$. The existence of partition $\mathcal Q$ is proved in lemma \ref{lem:parts-count}.
 \item  For each $Q$ in $\mathcal{Q}$, make $G'[Q]$ a clique. For each $v \in Q$, define $d_{G'}(v) = |Q|-1$.
 \item Let $A$ be a set consisting of exactly one vertex from each $Q$ in $\mathcal Q$, $|A| = N-2$. 
 \item Let $f(v)=d_{G'}(v)+2$ for each vertex $v$ in $A$ and $f(v)=d_{G'}(v)$ for each $v$ in  $G'\setminus A$.
 \item Let  $u_1$ be a vertex in $G$. 
 \item {For each 4-vertex path $u_0,u_1,u_2,u_3$ in $G$, do the following:
 \begin{enumerate}
 \item  Let $\sigma$ be a bijection from $V(G)\setminus\{u_1,u_2\}$ to $A$. 
 \item For each edge $\{u,v\}$ in $G\setminus\{u_1,u_2\}$, add edge $\{\sigma(u),\sigma(v)\}$ to $G'$.\textit{//Make $G\setminus \{u_1,u_2\}$ isomorphic to $G'[A]$ under $\sigma$} 
 \item Fix $f(\sigma(u_0))=d_{G'}(v)+1$ and $f(\sigma(u_3))=d_{G'}(v)+1$. \textit{//For a connected $f$-factor $H'$ in $G'$, the graph $H'[A]$ is a spanning path with  $sigma(u_0)$ and $sigma(u_3)$ as end points}
 \item Output $(G',f)$.
 \item \textit{//resetting $G'$ for the next iteration }
 \item Remove all the edges in $G'[A]$ from $G'$.
 \item Fix $f(\sigma(u_0))=d_{G'}(v)+2$ and $f(\sigma(u_3))=d_{G'}(v)+2$.
 \end{enumerate}}
\end{enumerate}
Observe that if the number of vertices $N$ in $G$ is sufficiently large (depending on $\epsilon$),  the total space required to hold $\mathcal G$ output by $\textsf R$ is  $2^{o(N)}$. 
\begin{lemma}
Let $m$ be an integer and let $k<\sqrt{m}$. Then $\floor{{m}/{(\ceil{m/k}+1})}\geq k-2$. 
\label{lem:parts-count}
\end{lemma}
\noindent
\textit{Proof in Appendix.} \\
The following lemma proves the correctness of the reduction.
\begin{lemma}
The graph $G$ has a Hamiltonian cycle $H$ if and only if $\textsf R$ outputs a pair $(G',f)$ such that $G'$ has a connected $f$-factor $H'$.
\label{lem:reductionlemma}
\end{lemma}
\noindent
\textit{Proof in Appendix.} \\
If we have a polynomial time algorithm for the connected $f$-factor problem for a given constant $\epsilon > 0$, then we test for the existence of a connected $f$-factor of $G'$ for each $(G',f)$ in $\mathcal G$. The size of the set $\mathcal G$ is $\mathcal O(N^3)$ where $N=|V(G)|$. Computation of each $G'$ takes $2^{o(N)}$ time. Thus, in time $2^{o(N)}$, we check for the existence of a pair $(G',f)$ in $\mathcal G$ such that $G'$ has a connected $f$-factor. 
\begin{lemma}
Let $\epsilon > 0$ and let $f(v)\geq n/g(n))$ for every $v$ in $V$ and $g(n)$ is in  $\mathcal O((\log{n})^{1+\epsilon})$.  Then the connected $f$-factor problem is not in \P 
unless the ETH is false.  
\label{lem:non-p}
\end{lemma}
From lemmas \ref{lem:non-npc} and \ref{lem:non-p}, we come up with the following theorem.
\begin{theorem}
Let $G$ be a graph having $n$ vertices and $f$ be a function where $f(v)\geq n/g(n)$ for each $v$ in $V$. For each $\epsilon > 0$ and each $g(n)$ in $\mathcal O((\log{n})^{1+\epsilon})$,  the connected $f$-factor problem is in \NPI\   ~unless the ETH is false.
\end{theorem}
%
%
\section{Properties of alternating circuits and $f$-factors}
\label{sec:ac}
To start with, we present properties of alternating circuits which we use extensively.  Alternating circuits are intricately related to $f$-factors as they provide a way of {\em moving} from one $f$-factor to another.  We present the following lemmas from our previous work in \cite{NR15} and the proofs of the lemmas in this section, which are necessary  are in the appendix.
\begin{lemma} \label{lem:rblemma}
Let $T$ be a graph in which each edge is assigned a color from the set $\{red,blue\}$. Each component in $T$ is an alternating circuit if and only if  $d_R(v)=d_{B}(v)$ for  every  $v$ in $T$. 
\end{lemma}
Consider two $f$-factors $H_1$ and $H_2$ of a graph $G$. If color the edges in $H_1$ with color red and those in $H_2$ with color blue, then each component in $H_1\bigtriangleup H_2$ is an alternating circuit. Note that if two alternating circuits $T_1$ and $T_2$ have a vertex in common, then $T_1\cup T_2$ is an alternating circuit. 
\begin{lemma} \label{lem:symdiff}
Let $H$ and $H'$ be two $f$-factors of $G$. If $T=H\bigtriangleup H'$ (symmetric difference of the edge sets) then $T$ is a switch on both $H$ and $H'$. 
\end{lemma}
%
\begin{lemma} \label{lem:minalterswitchlemma} 
Let $H$ be a subgraph of $G$ and let $T$ be a switch on $H$. Assign color red to edges in $T\cap H$ and blue to those in $T\setminus H$. If $T$ is a minimal alternating circuit and $G'=$Switching($H$,$T$), then $|N_{H}(v)\cap N_{G'}(v)| \geq d(v) -2$, for each $v $ in $ V$.
\end{lemma}
\begin{lemma}  \label{lem:enforcedg}
 Let $S \subseteq E(G)$. An $f$-factor $H$ containing all the edges in $S$, if one exists, can be computed in polynomial time. 
 \end{lemma}
{\bf Decomposing an alternating circuit into minimal alternating circuits.}
In our algorithm we repeatedly take an alternating circuit and decompose into a set of minimal alternating circuits containing a given set of edges.  
The function Min-AC-Set($U$,$S$) in ~\cite{NR15} take an alternating circuit $U$ and a set of edges $S \subseteq U$ as input and output a set $\mathbb U$ of edge disjoint minimal alternating circuits each of which is present in $U$. Further, each edge in $S$ is present in some minimal alternating circuit $C$ in $\mathbb U$.  Min-AC-Set($U$,$S$)  identifies  an alternating circuit $C$ having $d_R(v)$ and $d_B(v)$  at most 2 and adds it to $\mathbb U$ only if some edge in $S$ is present in $C$. Further it removes the  identified $C$ from $U$. This step is repeated until  $U$ is empty. The crucial step in Min-AC-Set($U$,$S$) is to find a minimal alternating circuit in $U$.  This is presented in the recursive Procedure Find-Min-AC($U$)(in Appendix).
\begin{lemma}
The procedure Min-AC-Set($U$,$S$) outputs a set $\mathbb U$ of edge disjoint minimal alternating circuits each of which has at least one edge from $S$.
\label{lem:min-ac-set}
\end{lemma}
\section{Algorithm for computing a Connected $f$-factor} \label{sec:con-f-factor}
In this section we complete the algorithm  outlined  in Section \ref{sec:overview}. The algorithm takes an unweighted graph $G$ and a function $f$ as input and  outputs a connected $f$-factor of $G$ if it exists. When the function $f(v)\geq n/g(n)$ for each $v$ in $V$ and $g(n)$ is polylogarithmic in $n$, the algorithm runs in time $\tilde{\mathcal O}(n^{2g(n)})$.  We start with a justification of why $g(n)$ being  polylogarithmic in $n$ is crucial for our analysis.  
\begin{lemma} 
\label{lem:whypolylog}
Let $g$ be a function on the set of positive integers. Let there be a positive constant $b$ such that for each positive integer $n$, $g(n) \leq b (\log n)^b$. Then for each $n \geq n_0$, $2(g(n))^4$ is at most $n$ for a sufficiently large constant $n_0$.
\end{lemma}
The proof of the above lemma is easy as $n/2b^4$ is asymptotically larger than $(\log n)^{4b}$ for any constant $b$. Lemma \ref{lem:whypolylog} is crucial in the analysis of our algorithm. Algorithm \ref{Algm:1} processes the input graph based on $n$. If $n$ is smaller than $n_0$, it exhaustively checks for a connected $f$-factor. If $n$ is at least $n_0$,
then by Lemma \ref{lem:whypolylog}, $n \geq 2(g(n))^4$ and we use this to bound the running time of our algorithm.\\
{\bf Description of the Algorithm}
The idea is to start with an arbitrary $f$-factor $H_0$ of $G$ and compute a connected $f$-factor using the template in Section \ref{sec:overview}. We use a recursive subroutine Restricted-$f$-Factor($H_{i-1}$,$\mathcal Q_{i-1}$) which returns a  connected $f$-factor if it exists or it returns an empty graph otherwise. 

\noindent
\begin{algorithm}[H]
\label{Algm:1}
\caption{The Algorithm for deciding connected $f$-factor when $g(n)$ is in $\mathcal O(\polylog{n})$}
\textit{Input:}{$G(V,E)$,$f$}\;
\textit{Output:}{$G'$, a connected $f$-factor  of $G$ if exists, and outputs failure otherwise.}\;
\If {$n \leq n_0$ }{ exhaustively search for a connected $f$-factor.\; 
Return the connected $f$-factor if it is found, else return failure.\;}
$H_0$=Tutte's-Reduction($G$,$f$). \;
\If {$H_0=empty$}{ then exit reporting failure.\;}
\If {$H_0$ is connected}{ \textit{Output} $H_0$ and exit.\;}
$\mathcal Q_0$=$\{V(G)\}$.
\Comment{Initialize the root partition $\mathcal Q_0$ with a single part $V(G)$}\;
$G'$=Restricted-$f$-Factor$(H_0, \mathcal Q_0)$.\;
\If {$G'$ is nonempty } {\textit{Output} $G'$ and exit.}{\textit{Output} failure}.
 \end{algorithm}
\noindent
The following lemma plays a critical role in the correctness of our subroutine  {Restricted-$f$-Factor($H_{i-1}$,$\mathcal Q_{i-1}$)}.
\begin{lemma}
Let $G$ be a graph having a connected $f$-factor. Let $\mathcal Q$ be  a partition of the vertex set $V$. There exists a spanning tree $T$ of $G/\mathcal Q$ and an $f$-factor $H$ of $G$ such that $E(T)\subseteq E(H)$. Further given $T$, $H$ can be computed in polynomial time.
\label{lem:connect-parts}
\end{lemma}
\begin{proof}
Let $G'$ be a connected $f$-factor of $G$. For any partition $\mathcal Q$ of the vertex set,  $G'/\mathcal Q$ is connected. Consider a spanning tree $T$ of $G'/\mathcal Q$.  Clearly, there exists at least one $f$-factor $H$ containing $E(T)$ and hence $H/\mathcal Q$ is connected. Once we have $E(T)$, $H$ can be computed in polynomial time using Lemma \ref{lem:enforcedg}. 
 \end{proof}
We now present the recursive procedure Restricted-$f$-Factor() which expands the outline in Section \ref{sec:overview}.

\noindent
\begin{algorithm}[H]
\caption{The procedure Restricted-$f$-Factor() recursively compute a connected $f$-factor of $G$, if it exists.}
  \SetKwFunction{proc}{Restricted-$f$-Factor}
 \SetKwProg{myproc}{Procedure}{}{}
  \myproc{\proc{$H_{i-1}$,$\mathcal Q_{i-1}$}}{
 
$\mathcal Q_i$=$empty$.\;
\For {each $X\in \mathcal Q_{i-1}$}{
$\mathcal Q_1$=$\mathcal Q_i \cup \{Y| Y$ is vertex set of a maximal component in $H[X]\}$.\;
} 

\If{$\mathcal Q_i=\mathcal Q_{i-1}$}{return $H_{i-1}$ and exit. \Comment{$H_{i-1}$ is a connected $f$-factor} \;}
$G'$=$empty$.\;
$H'_i$=$empty$.\;
\Comment{ BEGIN {\bf Partition-Connector}($\mathcal Q_i$)} \;
\For {each spanning tree $T$ of $G/\mathcal Q_i$}{ 
$T$=$T\setminus H_{i-1}$.\Comment{Ignore edges that are already there in $H_{i-1}$}\;
\If {an $f$-factor $H'_i$ containing $E(T)$ exists}{  exit  \textit{Loop 2}.\Comment{$steps$ $12\ldots 18$}}
}
\If{$H'_i$==$empty$}{exit and return empty.\Comment{There does not exist an $f$-factor that connects $\mathcal Q_i$}}
\Comment{ END {\bf Partition-Connector}($\mathcal Q_i$)}\;
\Comment{BEGIN {\bf Next-Factor}($H_{i-1}$,$H'_i$,$\mathcal Q_i$)}\;
$S$=$E(H_{i-1})\bigtriangleup E(H'_i)$.\;
$\mathcal S$=$empty$.\Comment{Set of minimal alternating circuits containing $E(T)$}\;
\For {each component $U\in S$}{ 
$\mathcal S$=$\mathcal S\cup $Min-AC-Set($U$,$T$)}
$H_i=$Switching($H_{i-1}$,$\mathcal S$). \Comment{$H_i$ is an $f$-factor containing $E(T)$}\;
\Comment{END {\bf Next-Factor}($H_{i-1}$,$H'_i$,$\mathcal Q_i$)}\;
$G'$=Restricted-$f$-Factor($H_i$,$\mathcal Q_i$).\; 
return $G'$.\;
}\end{algorithm}
\noindent
We use the following lemma in arguing the correctness of our algorithm.
\begin{lemma}
\label{lem:partsizebound}
If $G$ has at least $2(g(n))^4$ vertices, then in each recursive call to Restricted-$f$-Factor($H_{i-1}$,$\mathcal Q_{i-1}$), the number of parts in $\mathcal Q_{i-1}$ is at most $g(n)$. 
\end{lemma}
\begin{proof}
Consider the computation of $H_i$ from $H_{i-1}$ in the first call to the subroutine with parameters $H_0$ and $\mathcal Q_0$ from Algorithm 1. In each iteration of loop 2, the number of edges in $T$ is at most $g(n)-2$. This is because the number of components in $H_0$ is at most $g(n)-1$. Consider $\mathcal S$ computed in step 24. We color edges in $H_{i-1} \cap \mathcal S$ with color red and those in $H'_i\cap \mathcal S$ with color blue. From  Lemma \ref{lem:min-ac-set}, minimality of each $s$ in $\mathcal S$ computed in step 27 and Lemma \ref{lem:minalterswitchlemma}, $|N_{H_{i-1}}(v)\cap N_{H_i}(v)|$ is at least $n/g(n)-2(g(n)-2)$ for each vertex $v$ in $V$. 

\noindent
Assume that there exists a recursive call in which the number of parts in $\mathcal Q_{i-1}$ is more than $g(n)$. We prove that this contradicts our premise that $G$ has at least $2(g(n))^4$ vertices.   Let the pairs $(H_0,\mathcal Q_0), (H_1,\mathcal Q_1),\ldots,(H_k,\mathcal Q_k),\ldots$ be the sequence of arguments to Restricted-$f$-Factor($H_{i-1}$,$\mathcal Q_{i-1}$). Let the number of parts in $\mathcal Q_k$  be larger than $g(n)$ and for each $0 \leq i \leq k-1$, $|\mathcal Q_i| \leq g(n)$. As discussed above, $k$ can not be $0$.
\noindent
Observe that for each $1 \leq i \leq k$,  $|\mathcal Q_i| > |\mathcal Q_{i-1}|$ and $|\mathcal Q_0|=1$.   This implies $|\mathcal Q_{i}| \geq i+1$ for every $i$. Thus the level number $k-1$ is at most $g(n)-1$ and hence $k$ is at most $g(n)$. 
Let $T$ be a spanning tree in $G/\mathcal Q_i$ and let $H'_i$ be the $f$-factor in the $i$-th recursion that contains $T$(connects $\q_{i}$).  Let $S$ be the 
symmetric difference $E(H_{i-1})\bigtriangleup E(H'_i)$, and let $\mathcal S$ be the subset of decomposition of $S$ into minimal alternating circuits which we use for switching in step 29. For each $0\leq i \leq k-1$, $\mathcal Q_{i}$ has at most $g(n)$ parts. The number of parts in $\mathcal Q_{k-1}$ computed in the recursive call Restricted-$f$-Factor($H_{k-2}$,$\mathcal Q_{k-2}$) is at most $g(n)$. This implies in each of those recursive calls Restricted-$f$-Factor($H_i$,$\mathcal Q_i$) where $0\leq i \leq k-2$, the number of edges in $T$ computed in step 13 is at most $g(n)-1$.  Consequently, from Lemma \ref{lem:min-ac-set},
the number of  minimal alternating circuits in $\mathcal S$ is at most $g(n)-1$ for each recursive call with parameters ($H_i$,$\mathcal Q_i$) for $0\leq i < k-1$.   Thus, from Lemma \ref{lem:minalterswitchlemma},
the size of $\displaystyle \cap_{0 \leq i\leq k-1} N_{H_i}(v)$ is at least $n/g(n)-2(g(n)-1)(k-1)$. Further, $|N_{H_0}(v)\cap N_{H_{k-1}}(v)|\geq n/g(n)-2(g(n)-1)(k-1)$ where $H_{k-1}$ computed at the end of the call Restricted-$f$-Factor($H_{k-2}$,$\mathcal Q_{k-2}$). This means for each vertex $v$ in $V$, at least $n/g(n)-2(g(n)-1)(k-1)$ edges incident on $v$ in $H_{k-1}$ were also present in $H_0$. Since $k\leq g(n)$, we get the size of $N_{H_0}(v)\cap N_{H_{k-1}}(v)$ to be at least $n/g(n)-2(g(n)-1)(g(n)-1)$. Further, each part in  $\mathcal Q_{k}$ computed in the call Restricted-$f$-Factor($H_{k-1}$,$\mathcal Q_{k-1}$) has more than $n/g(n)-2(g(n)-1)(g(n)-1)$ vertices. This implies that the total number of vertices counted in the parts of  $\mathcal Q_k$ is more than $(g(n)+1)\cdot(n/g(n)-2(g(n)-1)(g(n)-1))$. Clearly, the total number of vertices $n$ should be larger than $\frac{g(n)+1}{g(n)}(n - 2 (g(n))^3)$.   By rearranging the terms we get $n < 2(g(n))^4$.  This contradicts our premise that $G$ has at least $2(g(n))^4$ vertices.  Therefore, our assumption that $|\mathcal Q_k| > g(n)$ is wrong. Hence the lemma.
 \end{proof}
\begin{lemma}
\label{lem:noofcalls}
Let $n$ be at least $2(g(n))^4$. The number of times the subroutine Restricted-$f$-Factor($H_{i-1}$,$\mathcal Q_{i-i}$) gets invoked recursively is at most $g(n)-1$. 
\end{lemma}
\begin{proof}
This follows immediately from Lemma \ref{lem:partsizebound}, as the number  of parts in $\mathcal Q_i$ is at most $g(n)$, and between two consecutive reclusive calls to  Restricted-$f$-Factor($H_{i-1}$,$\mathcal Q_{i-1}$), either a connected $f$-factor is found or we have a $\mathcal Q_i$ larger than $\mathcal Q_{i-1}$ to work with.
 \end{proof}
\noindent
Rest of the section contains the proof of the results discussed in Section \ref{sec:overview}.
\begin{proofof}{Running time in Theorem \ref{thm:ref-characterization-Algm1}}
If $G$ has an $f$-factor, then step 7 of Algorithm \ref{Algm:1} computes an arbitrary $f$-factor $H_0$.  The first call to Restricted-$f$-Factor($H_{i-1}$,$\mathcal Q_{i-1}$) is made with arguments $H_0$ and  $\mathcal Q_0 = \{V(G)\}$.  In the $i$-th  recursive call: if $\mathcal Q_i$ and $\mathcal Q_{i-1}$ are the same, then a connected $f$-factor is found, and this check can be done in polynomial time, and the correctness is by Fact \ref{fac:connect-parts}.  If there does not exist a $f$-factor $H'_{i}$ that connects $\mathcal Q_{i}$, then the algorithm exits at this step.  
This check is done by loop 2, and since the number of parts is upper bounded by $g(n)$, the time taken is at most $\binom{m}{g(n)-1}$, where $m$ is the number of edges in $G$.  This is the time taken to enumerate all spanning trees $T$ of $G/\mathcal Q_i$ and to check if there is a $f$-factor that contains the edges of $T$.  Thus Algorithm 1 completes in time $\tilde{\mathcal O}(n^{2g(n)})$.
If $G$ has a connected $f$-factor, then in at most $g(n)$-recursive calls, the Restricted-$f$-Factor($H_{i-1}$,$\mathcal Q_{i-1}$) will succeed due to Theorem \ref{thm:characterization}.   Further, by the same theorem, if $G$ does not have a connected $f$-factor, the procedure will terminate by identifying a partition $\mathcal Q$ that cannot be connected by an $f$-factor of $G$.   Hence the theorem.
  
\end{proofof}
\begin{proofof}{Characterization in Theorem \ref{thm:ref-characterization-Algm1}}
The forward direction of the proof is implied by Theorem \ref{thm:characterization} as a connected $f$-factor connects any partition, independent of the size of the partition. The reverse direction of the proof is from Algorithm \ref{Algm:1}. For each partition $\mathcal Q$ of size at most $g(n)$, if there exists an $f$-factor that connects $\mathcal{Q}$ then clearly the algorithm  computes a connected $f$-factor with in at most $g(n)-1$ recursions.
\end{proofof}

\section{Computing a minimum weighted connected $f$-factor}
\label{sec:weighted-factor}
\noindent
In this section we consider a variant where the input graph $G$ has positive weights on the edges, and the objective is to compute a minimum weighted connected $f$-factor. We consider the case where $f(v)\geq n/g(n)$ for some function $g(n)$ in $\mathcal O(\polylog{n})$. We extend Algorithm \ref{Algm:1} to solve this minimization problem as follows-
\begin{enumerate}
 \item In step 7 of Algorithm \ref{Algm:1}, instead of initializing $H_0$ with an arbitrary $f$-factor, we use { Modified-Tutte's-Reduction($G$,$f$)} to initialize $H_0$ with a minimum weighted $f$-factor of $G$.
 \item Further in loop 2 in Restricted-$f$-Factor($H_{i-1}$,$\mathcal Q_{i-1}$), $H'_i$ is the minimum weighted $f$-factor that connects $\mathcal Q_i$, if there exists one. 
\end{enumerate}
\noindent
It is clear, from the arguments in Theorem \ref{thm:ref-characterization-Algm1},  that these modifications still guarantee that the output will be a connected $f$-factor if one exists, and we have to show that the procedure computes an $f$-factor of minimum cost.  We refer to the above extension of Restricted-$f$-Factor($H_{i-1}$,$\mathcal Q_{i-1}$)  as Restricted-Min-$f$-Factor($H_{i-1}$,$\mathcal Q_{i-1}$).
\noindent
 To understand the behavior of the modification, recall that $H_0$ is a minimum weight $f$-factor of $G$. Secondly, let $H'_i$ be a  minimum weighted $f$-factor that connects $\q_{i}$  identified in the $i$-th recursion. The procedure builds $H_{i}$ from $H_{i-1}$ and $H'_i$. The  following lemma is used in bounding the cost of $H_i$ in the $i$-th recursive call  to Restricted-Min-$f$-Factor($H_{i-1}$,$\mathcal Q_{i-1}$).
\begin{theorem}
 Let $H$ be a minimum weighted $f$-factor of $G$ and let $T \subseteq E(G) \setminus E(H)$.  Let $H'$ be a minimum weighted $f$-factor among all $f$-factors containing  $T$. Let $S=E(H) \bigtriangleup E(H')$ and  let the edges of $S \cap H$ be colored red and the edges of $S \cap H'$ be colored blue.
Let $\mathcal S$ be a partition of $E(S)$ into minimal alternating circuits. The following are true:
\begin{enumerate}
\item For each $s $ in $ \mathcal S$, if $W(s)>0$ then $s \cap T \neq \phi$.
\item For any $\mathcal {S'} \subseteq \mathcal{S}$ satisfying $T\subseteq \displaystyle \cup_{s \in \mathcal {S'}}s$, Switching($H$,$\mathcal {T'}$) is an $f$-factor of weight exactly equal to $w(H')$.
\end{enumerate}
 \label{thm:uniqminac}
\end{theorem}
\noindent
The proof of Theorem \ref{thm:uniqminac}, which is also in our recent paper \cite{NR15}, is in the Appendix. The following lemma highlights an invariant which plays a critical role in arguing the correctness of the subroutine Restricted-Min-$f$-Factor().
\begin{lemma}
Consider the sequence of arguments $(H_0,\mathcal Q_0), (H_1,\mathcal Q_1),\ldots,(H_k,\mathcal Q_k)$ to the procedure Restricted-Min-$f$-Factor().  Let the input graph $G$ has a connected $f$-factor. The cost of $H_{k}$ is equal to that of $H'_{k}$ in the $k$-th recursive call to Restricted-Min-$f$-Factor().
\label{lem:weightBoundlemma}
\end{lemma}
\begin{proof}
For $k=1$, Theorem \ref{thm:uniqminac} directly completes the proof. For $k>1$, we prove by induction on the recursion level $i$. 
We assume the claim to be true for $k-1$ and we show this to be true for $k$.   We use $\mathcal S_i$ and $S_i$ to address the value of variables $\mathcal S$  and $S$ respectively, computed in step 28 of the function Restricted-Min-$f$-Factor($H_{i-1}$,$\mathcal Q_{i-1}$).  Assume that there exists a minimal alternating circuit $s$ in $H'_k\bigtriangleup H_k$ which is a switch on $H_k$ of negative weight. \\
We now derive  a contradiction to the optimality of $H'_{k-1}$. Note that $s$ is edge disjoint from $\bigcup_{s'\in \mathcal S_{k}}E(s')$. This is because,  $H'_k\bigtriangleup H_k$ is $S_k\setminus \mathcal \bigcup_{s'\in \mathcal S_{k}}E(s')$. Thus $s$ is a switch on Switching($H_{k}$,$\mathcal S_{k}$) of negative weight. Further, Switching($H_{k}$,$\mathcal S_{k}$) is $H_{k-1}$. 
By induction, $H_{k-1}$ is of weight $H'_{k-1}$ computed in the previous level of recursion. From the algorithm, Switching($H_{k-1}$,$s$) connects $\mathcal Q_k$. From the fact that $\mathcal Q_k$ is a refinement of $\mathcal Q_{k-1}$, it follows that Switching($H_{k-1}$,$s$)  is an $f$-factor that connects $\mathcal Q_{k-1}$ and is of weight less than
that of $H'_{k-1}$, and this contradicts the optimality of $H'_{k-1}$.
We now assume that there does not exist a  minimal alternating circuit $s$ of negative weight in $H_k\bigtriangleup H'_k$ which is a switch on $H_k$.  We show that such a switch $s$ of positive weight also cannot occur.   Note that such a positive weighted switch $s$ is a switch on $H'_k$ of negative cost. As described above, $s$ is edge disjoint from $\bigcup_{s'\in \mathcal S_{k}}E(s')$.  From the algorithm,  Switching($H'_k$,$s$) connects $\q_k$ and is of lower cost than $H'_{k}$.  Thus, We have a contradiction to the optimality of $H'_{k}$.   Therefore, the cost of $H_{k}$ is equal to that of $H'_{k}$ in the $k$-th recursive call to Restricted-Min-$f$-Factor().
 \end{proof}
\bibliography{biblio}
\textbf{\appendixname}
\ \\
\noindent
\begin{proofof}{Lemma \ref{lem:parts-count}}
Let $l=\ceil{m/k}+1$. Then $(l-2)k\leq m$. From $k<\sqrt{m}$, $(l-2)\geq (k-2)$. We have,
\begin{align*}
(l-2)k&\leq&m\\
(l-2)k -2(l-2)+2(k-2)&\leq&m\\
l(k-2)&\leq&m		
\end{align*}
\end{proofof}

\noindent
\begin{proofof}{Lemma  \ref{lem:reductionlemma}}
Consider the vertex $u_1$ selected in step 7 of $\textsf R$. Corresponding to each Hamiltonian cycle $H$ in $G$ there exists a unique 4-vertex path $u_0,u_1,u_2,u_3$ in $G$ and an associated iteration of step 8. Note that $H\setminus\{u_1,u_2\}$ is a path $L$ of length $N-2$ whose end vertices are $u_0$ and $u_3$. We have a corresponding spanning path $L'$ of length $N-2$ in $G'[A]$. Removing all the edges in $G'[A]\setminus L'$ from $G'[A]$ gives a connected $f$-factor $H'$ of $G'$. Conversely,  given a connected $f$-factor $H'$, we pick the edge $\{B^{-1}(u),B^{-1}(v)\}$ for each edge $\{u,v\}$ in $H'[{A}]$. This gives a path $L$ of length $N-2$ whose end vertices are some $u_0$ and $u_3$. From the reduction algorithm, there exists another path $u_0,u_1,u_2,u_3$ where $u_1$ and $u_2$ are the vertices that are not present in $V(L)$ and we have a Hamiltonian cycle in $G$.   
\end{proofof}
\noindent
\begin{proofof}{Lemma  \ref{lem:rblemma}}
In the forward direction, consider a component $C$ in $T$.  Since there is an Eulerian tour in $C$ in which consecutive edges are of different colors, it follows that $d_R(v)=d_{B}(v)$ for all  $v$ in $T$.  In the reverse direction, let $d_R(v)=d_{B}(v)$ for all  $v$ in $T$.  To complete the proof, we point to an exercise in  \cite[exercise 1.2.35]{DW00} which considers the formulation of Tucker's algorithm for computing an Eulerian circuit in \cite{AT76}.  First, at each vertex $v$ in a component $C$ in $T$, we pair each red edge incident on $v$ to a distinct blue edge incident on $v$. Since $d_{R}(v)=d_{B}(v)$, such a pairing is guaranteed to  exist. Secondly, using this pairing,  the required alternating circuit is the Eulerian circuit constructed by Tucker's algorithm.  Hence the lemma.
 \end{proofof}

\noindent
\begin{proofof}{Lemma  \ref{lem:minalterswitchlemma}}
Since $T$ is minimal, by definition, we know that $d_R(v)=d_B(v)$ for each $v $ in $ V$.  Consequently, not more than 2 edges incident on a vertex will be removed from $H$ as a result of applying Switching($H$,$T$).  Therefore, the number of common edges incident on $v$ in both $G'$ and $H$ is at least $d(v) -2$.
 \end{proofof}

\noindent
\begin{proofof}{Lemma  \ref{lem:enforcedg}}
Observe that removing the set of edges $S$ from an $f$-factor $H$ containing $S$, reduces the degree of each vertex $v$ in $H$ by $|\{(v,u) \in S\}|$. This is exactly an $f'$-factor of $G(V,E\setminus S)$ where $f'(v)=f(v)-|\{(v,u) \in S\}|$, for each $v $ in $ V$. Computing $f'$ and then  computing an $f'$-factor $H'$ of $G(V,E\setminus S)$ is easy. Recall that in polynomial time we can compute an $f'$-factor, if one exists, see West \cite{DW00}. Further adding the edges in $S$ to $H'$ gives an $f$-factor $H$ of $G$ containing $S$.
 \end{proofof}

\noindent
\begin{proofof}{Lemma  \ref{lem:connect-parts}}
Let $G'$ be a connected $f$-factor of $G$. For any partition $\mathcal Q$ of the vertex set, from Theorem \ref{thm:characterization}, $G'/\mathcal Q$ is connected. Consider a spanning tree $T$ of $G'/\mathcal Q$.  Clearly, there exists at least one $f$-factor $H$ containing $E(T)$ and hence $H/\mathcal Q$ is connected. Once we have $E(T)$, $H$ can be computed in polynomial time using Lemma \ref{lem:enforcedg}. 
 \end{proofof}

\noindent
\begin{proofof}{Theorem  \ref{thm:uniqminac}}
For any minimal alternating circuit $t$ which is a switch on $H$, 
recall that  $W(t)$, the weight of $t$, is $w($Switching($H$,$t$)$)-w(H)$. Since $H$ is optimum, for each  $t $ in $ \mathbb T$, $W(t)= w($Switching($H$,$t$)$)-w(H)\geq 0$.  Suppose there exists a minimal alternating circuit $t $ in $ \mathbb T$ such that $W(t) > 0$, and $t$ does not contain any of the edges in $S$.  Let us consider $T' = T \setminus t$, that is $T'$ is an alternating circuit obtained by removing the edges of $t$ from $T$, then $W(T')=W(T) - W(t)$.   Then Switching($H$,$T'$) is an $f$-factor containing $S$, and $w($Switching($H$,$T'$)$) = w(H) + W(T') = w(H) + W(T) - W(t) = w(H') - W(t) < w(H')$.   This contradicts the optimality of $H'$. Therefore, $ t \cap S \neq \phi$.  This implies for any subset $\mathbb {T'}\subseteq \mathbb T$ such that  $S\subseteq \displaystyle \bigcup_{t \in \mathbb {T'}}E(t)$, $w($Switching($H$,$\mathbb {T'}$)$)=w(H')$. 
 \end{proofof}

\begin{algorithm}[H]
\caption{The procedure Find-Min-AC($U$) returns a minimal alternating circuit in $U$}
  \SetKwFunction{proc}{Find-Min-AC}
 \SetKwProg{myproc}{Procedure}{}{}
  \myproc{\proc{$U$}}{
\If{$d_R(u)=d_B(u)\leq 2$ for each $u$ in $U$.}{Exit and return $U$.
{\Comment $U$ is a minimal alternating circuit}}\;
For each $u$ in $U$, pair each blue edge incident on $u$ to a distinct red edge incident on $u$.\;
Run Tucker's algorithm \cite[exercise 1.2.35]{DW00} on $U$ using the pairing defined in the previous step to get an Euler tour $\mathcal{T}$ in which consecutive edges are of different colors.\;
Let $v$ be a vertex with $d_R(v)>2$ in $U$.{\Comment Such a $v$ exists in $U$}\;
Start the tour $\mathcal T$ from $v$ and let $e_1$ be the edge through which $\mathcal{T}$ leaves $v$ for the first time and let $e_2$ be the edge through which $\mathcal{T}$ makes the first return to $v$. Let $e_3$ be the edge through which $\mathcal{T}$ continues the tour and let $e_4$ be the edge through which $\mathcal{T}$ makes the next return to $v$\; 
\If{$color(e_2)\neq color(e_1)$}{$U'=v,e_1,\ldots,e_2,v$.\;
\ElseIf{$color(e_4)\neq color(e_3)$}{$U'=v,e_3,\ldots,e_4,v$.\;
\Else{$U'=v,e_1,\ldots,e_4,v$.{\Comment $color(e_1)\neq color(e_4)$}\;}}}
Return Find-Min-AC($U'$).\;}
 \end{algorithm}

\end{document}